\documentclass{aa}  
\usepackage{graphicx}
\usepackage{txfonts}
\usepackage{natbib}
\usepackage{url}

\begin{document}

   \title{Mg\,I emission lines at 12\,\&\,18\,$\rm\mu$m in K giants\thanks{Partly based on observations 
obtained at the Gemini Observatory, which is operated by the 
Association of Universities for Research in Astronomy, Inc., under a cooperative 
agreement with the NSF on behalf of the Gemini partnership: the National Science 
Foundation (United States), the Science and Technology Facilities Council (United 
Kingdom), the National Research Council (Canada), CONICYT (Chile), the Australian 
Research Council (Australia), CNPq (Brazil) and SECYT (Argentina).}}
           
   \author{J.O. Sundqvist\inst{1}\and
           N. Ryde\inst{2,3}\and
           G.M. Harper\inst{4}\and
           A. Kruger\inst{5} \fnmsep \thanks{Visiting Astronomer at the Infrared Telescope Facility, which is operated by the 
           University of Hawaii under Cooperative Agreement no. NCC 5-538 with the National 
           Aeronautics and Space Administration, Science Mission Directorate, Planetary Astronomy 
           Program.}\and
           M.J. Richter\inst{5} \fnmsep $^{\star \star}$}
          
   \institute{Institut f\"ur Astronomie und Astrophysik der Universit\"at  
              M\"unchen, Scheinerstr. 1, 81679 M\"unchen, Germany\\
              \email{jon@usm.uni-muenchen.de}\and
             Department of Astronomy \& Space Physics, Uppsala University, Box 515, 751 20 Uppsala, Sweden\and
             Lund Observatory, Box 43, SE-221 00 Lund, Sweden\and
             Center for Astrophysics and Space Astronomy,Astrophysics Research Lab, 593 UCB, 
             University of Colorado, Boulder, CO 80309-0593, USA\and
             Department of Physics, University of California at Davis, CA 95616, USA
             }
   \date{Received 14/03/2008 / Accepted 22/05/2008}

 
  \abstract
   {The solar mid-infrared metallic emission lines have already been observed and analyzed
   well, and the formation scenario of the Mg\,I 12 $\rm\mu$m lines 
   has been known for more than a decade. 
   Detections of \textit{stellar} emission at 12 $\rm\mu$m have, however,
   been limited to Mg\,I in very few objects. Previous modeling 
   attempts have been made only for Procyon and two cool evolved stars, 
   with unsatisfactory results for the latter. This prevents 
   the lines' long predicted usage as probes of stellar magnetic fields.}
   {We want to explain our observed Mg\,I emission lines at 12 $\rm\mu$m 
in the K giants Pollux, Arcturus, and Aldebaran and at 18 $\rm\mu$m in Pollux
and Arcturus. We discuss our modeling of these lines and particularly  
how various aspects of the model atom affect the emergent line profiles.}
   {High-resolution observational spectra were obtained using TEXES at Gemini North and the IRTF. 
   To produce synthetic line spectra, we employed standard one-dimensional, 
plane-parallel, non-LTE modeling for trace elements in cool 
stellar atmospheres. We computed model atmospheres with the MARCS code, 
applied a comprehensive magnesium model atom, and used the radiative transfer code 
MULTI to solve for the magnesium occupation numbers in statistical equilibrium.} 
   {The Mg\,I emission lines at 12 $\rm \mu$m in the K giants are stronger than in the dwarfs 
observed so far. We present the first observed stellar emission lines from Mg\,I at 18 $\rm \mu$m
and from Al\,I, Si\,I, and presumably Ca\,I at 12 $\rm \mu$m.
We successfully reproduce the observed Mg\,I emission lines simultaneously in the giants and in the Sun,
but show how the computed line profiles depend critically on atomic data input and how the inclusion 
of energy levels with $n \ge 10$ and collisions with neutral hydrogen are necessary 
to obtain reasonable fits.}
   {} 


   \maketitle


\section{Introduction}
\label{Introduction}
Metallic solar emission lines around 12\,$\rm\mu$m were first identified by 
\citet{Chang83}, with the most prominent lines originating 
from transitions\footnote{Quantum state $nl$, 
where $n$ denotes the principal quantum number and $l$ 
the orbital} $7i~\rightarrow~6h$ 
(12.32\,$\rm\mu$m) and $7h~\rightarrow~6g$
(12.22\,$\rm\mu$m) between Rydberg states of neutral magnesium. 
Additional Rydberg emission lines from Al\,I,
Si\,I, and tentatively Ca\,I were identified as well \citep{Chang83,Chang84}. 
The Mg\,I line formation scenario remained unclear until \citet{Chang91} and 
\citet{Carlsson92}, hereafter C92, in two independent studies 
reproduced the emission features by employing standard plane-parallel 
numerical radiative transfer with a detailed atomic model and a reliable solar
atmosphere. They confirmed an origin below the atmospheric temperature 
minimum, refuted a chromospheric line contribution, and 
established a non-LTE (LTE: Local Thermodynamical Equilibrium)
formation scenario. 
The solar lines have subsequently been used in, e.g., Mg\,I 
statistical equilibrium analysis by Zhao et al.\ (1998).
C92 proposed a general Rydberg line formation 
mechanism for the highly excited metal lines, which 
implied that all visible metal emission lines in the solar spectrum 
around 12\,$\rm\mu$m originated in the photosphere. 
A detailed non-LTE modeling of the Al\,I emission has been 
carried out by \citet{Baumueller96}, where they confirmed this mechanism.\\ 
The lack of suitable spectrometers and the low stellar flux in the mid-infrared
have in the past made high-resolution spectroscopy in this 
wavelength region possible only for the 
Sun and a few luminous nearby stars. \citet{Ryde04} observed the 12\,$\rm\mu$m 
Mg\,I emission features in Procyon, and successfully reproduced the line profiles 
by employing the same modeling technique as C92. \citet{Uitenbroek96} 
observed and modeled the evolved stars 
Arcturus ($\alpha$ Boo) and Betelguese ($\alpha$ Ori). Using 
the same model atom as C92, they were unable to fit the line profiles of the 
$7i \rightarrow 6h$ Mg\,I transition, which appeared both in emission (Arcturus) and 
absorption (Betelguese). Their observational sample also included
five M giants and supergiants, in which the line appeared in absorption. 
However, \citet{Ryde06} investigated water vapor lines for Betelguese in the same spectral 
region, and found a water line that coincided with the wavelength of the Mg\,I 12.32\,$\rm\mu$m line.
The group successfully modeled the water line, without considering the Mg\,I blend
(which we predict to be very weak, see Sect.\ \ref{Observations_disc}). 
This may explain the sample of observed M star absorption at 12.32 $\rm \mu$m, 
since water vapor is expected in these stars, whereas the Mg\,I emission 
line contribution should be minor.\\
A well known potential use for the Mg\,I lines is as probes of 
magnetic fields, which play a fundamental 
role in the underlying physics of a cool stellar atmosphere.
Zeeman line-splitting from an external magnetic field increases quadratically with 
wavelength, while the Doppler broadening only has a linear dependence. 
Thus a line's sensitivity to magnetic fields becomes higher at longer wavelengths. 
The splitting of the solar emission lines was pointed out early and has been extensively 
analyzed. We have performed observations of 
the magnetically active dwarf $\epsilon$ Eridani. These will be reported on 
in a forthcoming paper (Richter et al., in preparation), hence we defer 
further discussions about stellar disk-averaged magnetic fields until then.\\
Prior to (stellar) diagnostic applications, however, we should make sure that we are able 
to model and understand these lines in a range of stars. So far, as mentioned above, 
modeling attempts for evolved stars have been unsuccessful. We address 
this issue here by analyzing high-resolution observational spectra, which 
show strong Mg\,I emission lines in the three giants 
Pollux (K0\,III), Arcturus (K1.5\,III), 
and Aldebaran (K5\,III). We model and analyze simultaneously 
the three K giants and the Sun, with particular emphasis on
influences from atomic data, and discuss why previous 
modeling attempts have not succeeded. The organization of the paper 
is as follows; in Sect.\ \ref{Observations}, 
we describe the observations. In Sect.\ \ref{Departure} we review some 
concepts about the formation of the infrared Mg\,I emission lines and 
in Sect.\ \ref{Modeling} we describe our modeling procedure. Results
are presented in Sect.\ \ref{Results} and we discuss them and give 
our conclusions in Sect.\ \ref{Discussion}. 


\section{Observations}
\label{Observations}

The observations were made with TEXES, the Texas Echelon-cross-echelle Spectrograph, 
\citet{Lacy02}. TEXES provides high spectral resolution in the mid-infrared and is 
available as a visiting instrument at both Gemini North and at IRTF, the Infra-Red 
Telescope Facility. The Pollux ($\beta$ Gem)
observations come from the November 2006 observing campaign at Gemini North. The Arcturus
and Aldebaran ($\alpha$ Tau) observations were done over many years at the IRTF. 
In most cases, the observations  
were primarily intended for flux calibration or focus tests and not to study 
the stars themselves.\\
When observing stars with TEXES, we nod the source along the slit, typically 
every 10 seconds, to remove sky and telescope background. Before each set of 8 to 16 nod 
pairs, we observe a calibration sequence that includes an ambient temperature blackbody 
and an observation of blank sky emission. The difference of blackbody minus sky serves 
as a first order telluric correction and flatfield. Where possible, a featureless 
continuum object with emission stronger than the target is also observed to further 
correct for telluric features and flatfielding. The largest asteroids work very well for 
this purpose, as does Sirius ($\alpha$ CMa) with respect to Pollux.\\
At the frequencies of the mid-infrared Mg I emission lines, the spectral orders from the TEXES high-resolution 
echelon grating are larger than the $256^2$ pixel detector array. This results 
in slight gaps in the spectral coverage. For the Pollux observations, which were the 
final observations before sunrise, we observed in two settings and adjusted the tilt 
of the collimator mirror feeding the echelon grating for the second setting. This shifts 
the spectral orders in the dispersion direction. By combining the data from these 
separate observational settings, we were able to fill in the gaps in the spectral orders. 
The Arcturus and Aldebaran data were constructed from many separate observing settings 
and no particular efforts were made to fill in the gaps.\\
Data reduction was done using a custom FORTRAN pipeline \citep{Lacy02}. The pipeline 
corrects for spikes and optical distortions in the instrument, allows the user to set the 
wavelength scale based on telluric atmospheric features, flatfields the data, differences 
nod pairs to remove the background emission, and then combines the resulting differences.
Finally it extracts a spectrum based on the spatial information within the two-dimensional
echellogram. The pipeline also provides a fairly accurate estimate of the relative noise 
in each pixel.\\
To combine data from separate observations, we first established a common wavelength scale. 
We corrected each spectrum for the Earth's motion at the time of the observation and then 
interpolated the data onto the common scale. We used a fourth-order polynomial derived 
from line free regions to normalize each spectral order. We determined the 
signal-to-noise ($S/N$) for the normalized spectrum via a Gaussian fit to pixel values and used 
the relative noise estimate established during pipeline reduction to assign a weight for 
each spectral pixel. When combining data, we choose to weight by the signal-to-noise 
squared, which effectively means weighting by successful observing time.\\
Observations of low pressure gas cells near 13.7 $\rm\mu$m at the November 2006 run 
indicate that the instrumental profile for the 12 $\rm \mu$m observations of 
Pollux has a Gaussian core with a FWHM $\sim 3.0\, \rm km\,s^{-1}$, corresponding to 
a spectral resolution $R \sim 10^5$. As the Arcturus and Aldebaran spectra 
combined data from four different runs and possible 
errors from the combinations may be significant, we were unable to make a reliable measurement 
of the instrumental profile in this region for these stars. At 18 $\rm \mu$m, 
similar measurements indicate that these observations have a Gaussian instrumental profile 
with a FWHM $\sim 4.5\, \rm km\,s^{-1}$. 
In Sect.\ \ref{Results} we display our observed data re-binned to approximately the 
spectral resolution, except for Fig.\ \ref{Fig:12um_obs}, where the pixel scale is used.
Signal-to-noise ratios in the spectra vary but are generally high, 
reaching $S/N \sim 450$ per pixel for Pollux and $\sim 300$ for Arcturus and Aldebaran, 
in regions around the 12.22 $\rm\mu$m line. At 18.83 $\rm\mu$m, the ratio is $S/N \sim 40$. 
  

\section{Departure coefficient ratios}
\label{Departure}
Before proceeding to a modeling description, we briefly 
review some important concepts about the formation of 
the 12\,$\rm\mu$m emission lines. In the following we use the
departure coefficients $b_i=n_i/n_i^*$,  
where $n_i$ is the actual number density (not to be confused 
with the principal quantum number $n$) of energy level $i$ 
and $n_i^*$ the corresponding LTE population, as calculated from 
the total magnesium abundance using the complete Saha-Boltzmann relations. 
In a spectral line, a departure coefficient ratio which differs from 
unity, $b_l/b_u \ne 1$, at line-forming depths 
causes a deviation of the line source function, $S_{\nu}^l$, from the 
Planck function, $B_{\nu}$, which affects the emergent intensity: 

\begin{equation}
\label{Eq:S/B}
\frac{S^l_{\nu}}{B_{\nu}} = \frac{e^{h\nu/kT}-1}{b_l/b_u \times e^{h\nu/kT}-1}
\end{equation}

\begin{figure}
                \resizebox{\hsize}{!}{\includegraphics[angle=90]{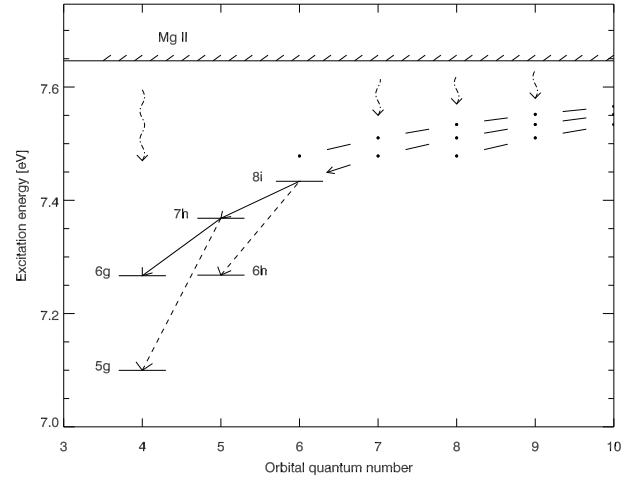}}
                \caption{Illustration of the `Rydberg ladder' (see text), 
                using a \textit{selected} part of the Mg\,I term diagram. 
                Five levels about the $7h \rightarrow 6g$
                transition are labeled with corresponding quantum numbers 
                $nl$. Dots mark energy levels with $n \geq 9$ 
                and $l \geq n-3$. Solid lines between levels show transitions 
                with $\Delta n, \Delta l = -1$, dashed show two 
                alternative transitions ($\Delta n = 2, \Delta l = -1$). 
                Dashed-dotted illustrate 
                recombination from the Mg\,II continuum.}               
        \label{Fig:terms}
\end{figure}

\noindent For a characteristic wavelength $\lambda = 12.3\,\rm\mu$m, and temperature 
$T=5000\,\rm K$, we get $e^{h\nu/kT}\sim 1.26$, and may directly from Eq.\ \ref{Eq:S/B} realize 
that already a small deviation from unity in the departure coefficient ratio 
causes a significant change in the line source function.
The physical reason for this is the increasing importance of stimulated emission in the infrared. 
If $b_u/b_l > 1$ and increases outwards in the atmosphere, we may get a rising total source function 
and a line profile appearing in emission despite an outwards decreasing temperature 
structure. Such departure coefficient divergence occurs between highly excited Rydberg levels 
in the outer layers of the modeled stellar photospheres considered in this study, and 
is the reason for the modeled emission lines.\\
Departure coefficient ratios that deviate from unity are set up by three-body 
recombination from the Mg\,II ground state and a `deexcitation ladder' that 
preferably takes $\Delta n,\Delta l=-1$ downward steps
(see Fig.\ \ref{Fig:terms} for an illustration).
In the solar case, all Mg\,I Rydberg levels 
are strongly collisionally coupled to each other and 
to the Mg\,II ground state. The main effects 
that drive the line source function out of LTE 
come from lines elsewhere in the term diagram, 
primarily lines between levels with intermediate 
excitation energies, which 
are optically thin in the outer atmosphere 
and experience photon losses \citep{Rutten94}. 
These levels impose a lower limit to the Rydberg state 
deexcitation ladder. The number densities of the 
Rydberg energy levels adjust to the upper and lower limits, 
and a radiative-collisional population flow occurs. 
It was shown, for the solar case by C92, how a 
high probability for $\Delta n,\Delta l=-1$ downward transitions is 
necessary for the Rydberg state ladder to be efficient and that 
these transitions dominate only if the collisional coupling 
is strong in the uppermost Mg\,I levels. It was also shown how 
this high probability arises from the regular character of the 
collisional cross-sections of transitions between highly excited levels. 
We thus remind of the remark in C92 that for a correct description of the 
ladder flow between highly excited levels, it is more important to have a 
\textit{consistent} set of collisional data, than to have the most accurate cross-sections 
for a few transitions.
These considerations are important to keep in mind when we later discuss our extension of the model atom
and the influence from collisions with neutral hydrogen.
A more comprehensive description of non-LTE effects 
throughout the Mg\,I term diagram that affect the 
solar 12\,$\rm\mu$m lines can be found in \citet{Rutten94}.

\section{Modeling}
\label{Modeling}
To produce synthetic line spectra, we employ standard one-dimensional, 
plane-parallel, non-LTE modeling for trace elements in cool 
stellar atmospheres. 
We generate model atmospheres from the MARCS code
\citep{Gustafsson75, Gustafsson08}, 
adopt a comprehensive magnesium model atom, and use the 
radiative transfer code MULTI \citep{Carlsson86, Carlsson92MULTI} 
to solve for the magnesium occupation numbers in statistical equilibrium,
while holding the structure of the atmosphere fixed. 

\subsection{Model atmospheres and stellar parameters}
\label{atmospheres}

The MARCS hydrostatic, plane-parallel models are computed on 
the assumptions of LTE, chemical equilibrium, homogeneity, and 
the conservation of the total flux (radiative plus convective; 
the latter treated using the mixing-length theory).
No chromospheric temperature rise is invoked but, as 
shown by C92, omitting a chromosphere has a negligible impact
on the solar Mg\,I 12 $\rm\mu$m transitions. The findings in this work 
and that by \citet{Ryde04} suggest that this holds true also for
other investigated stars. 
Apart from the temperature and density stratifications, 
a detailed MARCS radiation field was generated using opacity samplings 
including millions of lines. A sampled version of this radiation field was used when 
MULTI calculated the photoionization rates, in order to properly 
account for the line-blocking effect.  
We discuss some variations to our procedure in Sect. \ref{Discussion}.\\ 
The stellar parameters we used for Pollux were $T_{\rm eff}=4\,865 \rm\,K$, $\log g = 2.75$ (cgs), 
a solar metallicity (as given by \citeauthor{Grevesse07} 2007), 
and a depth independent `microturbulence' $\xi =1.5\,\rm km\,s^{-1}$, 
all based on a spectral analysis of optical iron and calcium lines made by \citet{Drake91}. 
For the parameters of Aldebaran we adopted $T_{\rm eff}=3\,900 \rm\,K$, $\log g = 1.5$ (cgs), 
a metallicity\footnote{Here
we do not specify individual metal abundances, but [M/H] is taken 
as the characteristic metallicity where as usual
[A/B] = $\rm \log(n_A/n_B)_* - \log(n_A/n_B)_{\sun}$.} [M/H]\,$=-0.25$, and $\xi =1.7\,\rm km\,s^{-1}$. 
These are from primarily \citet{Decin03} but considering 
also sources accessible at the SIMBAD astronomical database.
Finally for Arcturus we used $T_{\rm eff}=4\,280 \rm\,K$, $\log g = 1.5$ (cgs),
[M/H]\,$=-0.50$, and $\xi =1.7\,\rm km\,s^{-1}$.
A discussion of these Arcturus parameters can be found in \citet{Ryde02}.
The stars are all nearby and well-studied objects, and their 
parameters should be fairly accurate. Model grids show that the effects on the lines 
from (reasonable) variations in $\log g$ or $T_{\rm eff}$ are 
smaller than effects from, e.g., atomic input 
data, which will be investigated in the following sections.\\
We convolved our computed intrinsic line profiles with the instrumental profile, 
the projected rotational velocity ($v\,\sin i$), and the `macroturbulence' 
(none of which affect the line strength but only the profile shape). 
We adopted $v\,\sin i$ values from \citet{Smith79}, which for 
Pollux, Arcturus and Aldebaran are, respectively, $v\,\sin i = 0.8,\,2.7,\,2.7\,\rm km\,s^{-1}$. 
As we were unable to obtain a fair estimate of the instrumental profile for the
Arcturus and Aldebaran spectra around 12 $\rm \mu$m, we choose first to assign an isotropic Gaussian shape 
with characteristic Doppler velocity $v_m$ for the combined effect of the instrumental profile and the macroturbulence.
For Pollux, where the instrumental profile could be separated out, 
we obtained $v_{macro}$\,$\sim$\,3.3\,$\rm km\,s^{-1}$. 
However, it became clear that the modeled line wings of the K giants better 
fitted the observations when assuming a radial-tangential Gaussian shape \citep{Gray76B} 
for the macroturbulence. Therefore we decided to assign the Pollux 
instrumental profile for all three stars (the exact values are not 
so significant since the macroturbulence is the dominating external line broadening), 
and adopted a radial-tangential macroturbulence $v_{m,R-T} = 5.5,\,6.0,\,5.5\, \rm km\,s^{-1}$
to fit the observed line-widths. Our values are $\sim$\,2\,$\rm km\,s^{-1}$ higher than those measured 
from Fourier analysis in optical spectra by \citet{Smith79}.

\subsection{The model atom}
\label{Atom}
Our Mg\,I model atom is essentially an enlarged and slightly modified 
version of the one compiled by C92, and 
a full description can be found there. The original model atom 
has also been used in the analysis of solar magnetic 
fields (e.g., \citeauthor{Bruls95} 1995), 
for the Mg\,I 12\,$\rm\mu$m flux profiles of Procyon \citep{Ryde04},
and in a previous attempt to model 
giants \citep{Uitenbroek96}. In short, the atom is complete with all 
allowed transitions up to principal quantum number $n=9$ and 
includes the ground state of Mg\,II. 
We now describe changes and tests we have made.

\subsubsection{Enlargement of the model atom} 
\label{Enlargement}
\begin{figure}
                \resizebox{\hsize}{!}{\includegraphics[angle=90]{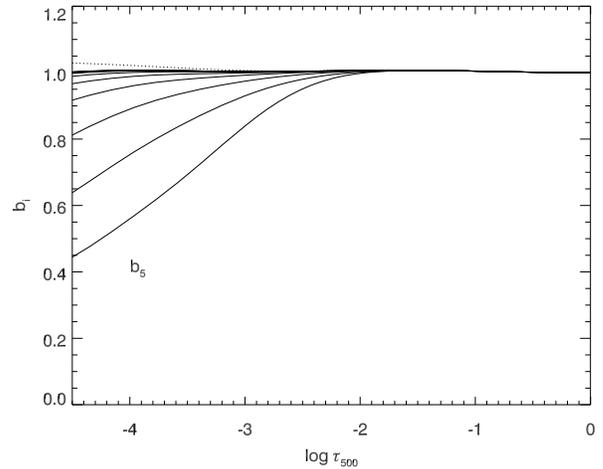}}
                \caption{Solar departure coefficients for Rydberg 
                state energy levels with $n\ge5$, as functions of the continuum 
                optical depth at 500 nm. $b_5$ is indicated in the figure, and an 
                increment of one follows upwards in the figure. The displayed 
                coefficients combine all $b_{n,l}$ coefficients with the same n and $l \geq 4$. 
                The Mg\,II ground state is labeled with dots.}
                \label{Fig:bs_Sun}
\end{figure}

\citet{Lemke87}, who also pointed out a possible photospheric 
line-origin through a rising line source function, 
made a statistical equilibrium investigation for the Sun 
but were unable to reproduce the Mg\,I 12\,$\rm\mu$m emission 
due to a combination of their adopted collisional data 
treatment and an inadequate model atom. 
Their exclusion of levels higher than $n=7$ resulted in an incorrect 
description of the replenishment of the Rydberg levels from the ion state. 
C92 experimented with the $n=8,9$ levels
and confirmed that these were necessary to have top-levels 
that were fully dominated by collisions and to obtain sufficient departure 
coefficient differences in line-forming layers 
to match the observed solar emission features. 
In this work we have extended the model atom to include levels 
with $n \ge 10$ to investigate if the departure coefficient 
differences are further enhanced.\\
The model atom was first enlarged to include all 
energy levels and allowed transitions with $n=10$.
To ensure homogeneity throughout the model atom, all new atomic data were calculated using the 
same formalisms as those employed by C92. The only exception was absorption oscillator strengths 
for transitions with $l \leq 3$ for which data was drawn from the opacity project (OP) TOP-BASE \citep{Cunto93}, 
since the tabulation used by C92 \citep{Moccia88} only extend to $n=9$. 
The enlargement caused an upward shift in the Mg\,I departure coefficients,
and the effect became more pronounced as $n$ increased; thus 
producing larger departure coefficient differences between 
adjacent levels. The same effect was seen 
in the Mg\,I statistical equilibrium for all our template atmospheres.
To investigate the influence of the low $l$ levels, 
a test-run was also made where only $n=10$ levels and transitions with 
$l \geq 4$ were included. This model atom and the complete $n=10$ atom 
produced almost indistinguishable results.\\ 
After this initial enlargement, atomic models were constructed step-wise, including 
higher principal quantum numbers. The enhancement continued until
the atom's uppermost $n$ level and the second uppermost were Boltzmann populated 
with respect to each other at all atmospheric layers (i.e., $b_{top-1}/b_{top}=1$). 
This criterion was met for all atmospheric models 
when reaching $n=15$, illustrated for 
the solar case in Fig.\ \ref{Fig:bs_Sun} (where our model 
actually meets the criterion already at $n=12$). 
Sensitivity tests verified that no differences in results 
occurred when adding the final $n=15$ top-levels. 
The main difference between the solar departure 
coefficients (Fig.\ \ref{Fig:bs_Sun})
and those of the K giants is that the latter have Mg\,II 
ground states that are more overpopulated relative to LTE in their outer atmospheres. 
This is mainly because Mg\,I and Mg\,II are 
competing ionization states in these cooler atmospheres, so that the ion
ground state becomes more sensitive to deviations from LTE in Mg\,I population densities. In 
Arcturus and Aldebaran, the overionization is further enhanced by the lower metallicity, 
which reduces the important line-blocking effect.\\ 
As the influence from levels with low $l$ was negligible already for $n=10$, we have confined 
the enlargement to levels and transitions with $l \ge 4$.

\subsubsection{Collisional data}
\label{Coll}
\citet{Ryde04} compared $radiative$ bound-bound and bound-free data from 
our model atom with OP data and found an overall good agreement. 
We have therefore restricted our analysis here to 
some aspects of the $collisional$ data, 
which usually introduce the largest 
uncertainty in the model atom, due to a 
large number of poorly known cross-sections.

\subsubsection{Collisions with neutral hydrogen}
\label{Hcoll}
The role of collisions with neutral hydrogen in cool stellar atmospheres 
has long been a subject for debate. Despite
small cross-sections (as compared to electron impacts), 
one may expect them to contribute significantly to 
collisional rates due to large $n_{HI}/n_e$ ratios. 
In the outer parts of the model 
atmospheres in this study, this ratio ranges from 
about $10^4$ in the Sun to $10^5$ in the cooler and 
more metal-poor Arcturus. No collisions with neutral 
hydrogen were considered in the original model atom.\\ 
When inelastic collisions with neutral hydrogen 
are explicitly included in non-LTE calculations,
a standard procedure is to adopt the 
recipe of \citet{Drawin69}, as given by \citet{Steenbock84}. 
The Drawin formula has often been criticized. 
\citeauthor{Steenbock84} state an accuracy of an
order of magnitude, but a rather common remark 
is that the recipe may overestimate the cross-sections 
with as much as one to six orders of magnitudes 
(see, e.g., \citeauthor{Asplund05} 2005, and references therein). 
Unfortunately, more reliable cross-sections are
scarce, especially for non-LTE calculations that 
require data for a large set of transitions.
A customary way around this shortage
is to adopt a scaling factor $S_{\rm H}$ to the Drawin formula,
calibrated on solar or stellar observations.\\
In studies concerning the solar Mg\,I 12 $\rm\mu$m lines, 
the Drawin formula was adopted by \citet{Lemke87} and \citet{Zhao98}.
The latter group scale their 
values with a factor that decreases exponentially with 
increasing excitation energy. Consequently, they apply
$S_{\rm H} =3\times10^{-10}$ for the 12 $\rm\mu$m 
transitions, which give them essentially 
the same result as if neglecting hydrogen collisions.
Recall also that the former group was unsuccessful in producing 
an emission line core in their statistical equilibrium analysis.\\
Here we have estimated the collisional rates due to neutral hydrogen 
impacts using the Drawin formula. When introduced without scaling factor, our models reveal 
solar 12 $\rm\mu$m lines in pure absorption and this case will therefore not be considered. 
By calibrating on the solar observations, we have adopted $S_{\rm H}=10^{-3}$
in all computations involving collisions with neutral hydrogen.  
However, in view of the existing uncertainties, we present results 
both from including these collisions for all radiatively allowed bound-bound 
(for which we have oscillator strengths) 
and all bound-free transitions, and excluding them. 

\subsubsection{Collisional excitation from electrons}
\label{Ecoll}
The bulk of the collisional cross-sections from electron impacts
for radiatively allowed transitions are calculated, as 
was done in C92, using the impact parameter approximation 
\citep{Seaton62}. \citet{Mashonkina96} showed that, overall,
this approach predicts significantly 
smaller cross-sections than the alternative semi-empirical formula of 
\citet{VanRegemorter62} (applied for the solar  Mg\,I
lines in, e.g., \citeauthor{Zhao98} 1998). \citet{Avrett94}
used both formalisms when modeling the solar Mg\,I lines 
and concluded that using \citeauthor{VanRegemorter62} gave somewhat weaker emission. 
In C92, the impact parameter approximation is claimed to give a consistent 
set of rates accurate to within a factor of two for transitions between 
closely spaced levels.\\
Note that the above mentioned formalisms relate the collisional cross-section 
to the oscillator strength and may therefore not be applied to radiatively forbidden transitions.
The `forbidden' cross-sections are here set to a multiplying factor times 
that of the closest allowed (see C92). The original choice (C92) for 
this factor was 0.05, but \citet{Bruls95} discovered some 
errors regarding a few oscillator strengths, 
accounted for here as well, and the factor was revised to 0.3 in order to reproduce the previous results.
We also adopt 0.3, which was used by \citet{Ryde04} as well.
A similar treatment for solar Mg\,I analysis 
has been used by \citet{Mauas88} who assumed 0.1, 
the same value as estimated in \citet{Allen73}.
\citet{Sigut96} also adopted 0.1 for Mg\,II (in work where they, 
for B type stellar photospheres, predicted the corresponding Rydberg emission
lines for Mg\,II), 
which they found to be in rough agreement with a few more rigorously calculated 
rates from low excitation transitions. We have tested 
using the enlarged model atom without collisions with neutral 
hydrogen and concluded that by a raise to 0.7 
times the cross-section of the nearest allowed transition
we are able to reproduce the observed solar lines, but that 
the modeled emission in the K giants remain far lower 
than the observed.\\
An alternative approach for the radiatively forbidden transitions,
applied in, e.g., Mg\,I (and II) non-LTE abundance analyses
\citep{Zhao98,Przybilla01,Gehren04,Mashonkina08}, is to 
set a constant collisional strength $\Omega =1$. 
Overall, this gives considerably lower 
collisional rates.
The collisional strength $\Omega$ for collisions with electrons
is related to the Maxwellian averaged downward 
collisional rate $C_{ji}$ [$\rm s^{-1}$] via:

\begin{equation}
C_{ji} = 8.63 \times 10^{-6} \Omega T_e^{-1/2}g_j^{-1}n_e
\end{equation}

\noindent where $T_e$ is the electron temperature,
$g_j$ the statistical weight of the upper level, and $n_e$ 
the number density of free electrons. The upward rate, $C_{ij}$, 
then follows from the principle of detailed balance. 
We have also tested assuming a constant $\Omega = 1$ on the solar and Arcturus model, 
and verified that the modeled emission for Arcturus still is inadequate,
remaining on the same low level as that displayed in Fig.\ \ref{Fig:12.22um_ArAl},
when collisions with neutral hydrogen are not included. The solar emission 
increases by this approach, but not enough 
to drastically change the suitable $S_{\rm H}$ factor. 
Overall, our experiments with the 
radiatively forbidden transitions tell us that the solar emission 
lines are somewhat sensitive to these rates, whereas the 
models of the K giants are much less responsive. 

\subsubsection{l-changing collisions}
\label{Lcoll}
No explicit calculations of collisional transitions of 
type \textit{n,l~$\rightarrow$~n,l'}
with high orbital quantum numbers ($l' \geq 4$) were made in
the original model atom, but rather it was assumed that these 
collisional rates were high enough to ensure strong coupling between 
closely spaced levels. This was established by setting $\Omega = 10^5$ 
if the difference in effective principal quantum number 
($n-\delta$ where $\delta$ is the quantum defect) in the transition was less 
than 0.1. The high value of $\Omega$ causes very high collisional 
rates, which essentially force all levels with the same 
$n$ and $l \geq 4$ to share a common departure coefficient
(i.e., to be Boltzmann populated with respect to each other), 
in agreement with the proposed assumption. 
This, however, only holds for all relevant atmospheric layers in the Sun, 
whereas in the more diluted model atmospheres of the K giants, we find 
deviations between departure coefficients in outer layers (for Arcturus 
outside $\log \tau_{500} \sim -2$) for levels with
relatively low orbital quantum numbers, when setting $\Omega =10^5$. 
Apparently, such a value does not suffice for the K giants and
we need to either raise the factor or explicitly 
estimate the rates. Note that C92 based the 
assumption of common departure coefficients 
mostly on the large cross-sections for $l$-changing collisions with neutral 
hydrogen calculated by \citet{Omont77}, cross-sections later shown to be 
over-estimated by an order of magnitude \citep{Hoangbinh95}.\\
In this work, we have calculated explicit electron/ion collisional rates 
for transitions with $\Delta l= \pm 1,\Delta n=0$ and 
$l,l' \geq 3$ by using a cut-off at large impact parameters, 
as outlined by \citet{Pengelly64}. We find that the radial cut-off for 
transitions with low $l$ is set by the non-degeneracy of the energy levels.
For ion rates, this happens inside the
radius where the strong interaction dominates and
hence the impact parameter approximation is not expected to be reliable.
However, for electron impacts the cut-off is at larger distances 
and since electron rates dominate over ion rates for these 
transitions, our approach should provide rates accurate to 
at least an order of magnitude also in the low $l$ range. 
Ion rates surpass electron rates from $l$\,$\sim$\,9 
and higher, where the cut-off is
well within the weak-interaction limit. We assume 
only singly ionized elements ($n_{ion} = n_e$) 
with the largest electron/ion donor being
magnesium, providing $\sim$\,40\,\% of the total 
electron/ion pool in the relevant atmospheric layers
in our MARCS models.\\ 
The calculated rates are in good agreement with 
the electron rates tabulated in \citet{HoangBinh94}, who 
exclusively considered $l$-changing collisions 
for the $n=6,7$ levels. Our rates are 
higher than those inferred from 
$\Omega = 10^5$ (by typically a factor of $\sim$\,4 
for, e.g., $n=7$), hence no changes in results occur
for the solar atmosphere. For the K giants, 
small differences between departure 
coefficients with the same $n$ and $l \geq 4$ still exist but 
effects from using explicit rates are small. When forcing 
common departure coefficients for all equal $n$ levels with $l \geq 4$
(by drastically increasing our computed rates), we still obtain 
normalized emission peaks for the K giants
that differ only by a few percentage points.


 \section{Results}
\label{Results}

\begin{figure}
                \resizebox{\hsize}{!}{\includegraphics[angle=90]{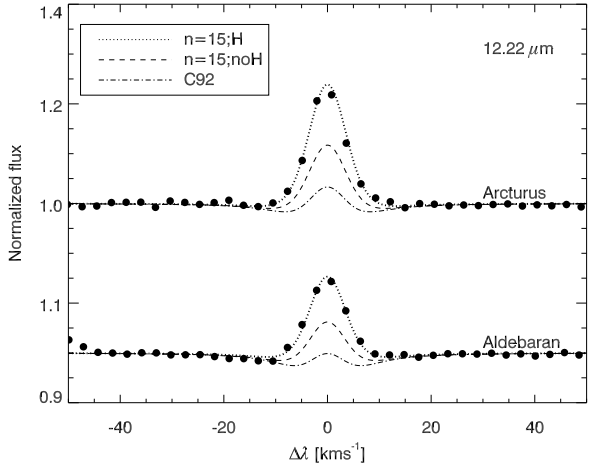}}
                \caption{Observed and modeled line profiles in Arcturus 
                and Aldebaran for the 12.22 $\rm\mu$m line, plotted 
                on a velocity scale. Labels as indicated in the figure; 
                where C92 denotes the original model atom, n=15;noH the 
                extended excluding collisions with neutral hydrogen and 
                n=15;H the one including such. The filled dots denote the 
                observed data.}
        \label{Fig:12.22um_ArAl}
\end{figure} 

\begin{figure}
               \resizebox{\hsize}{!}{\includegraphics[angle=90]{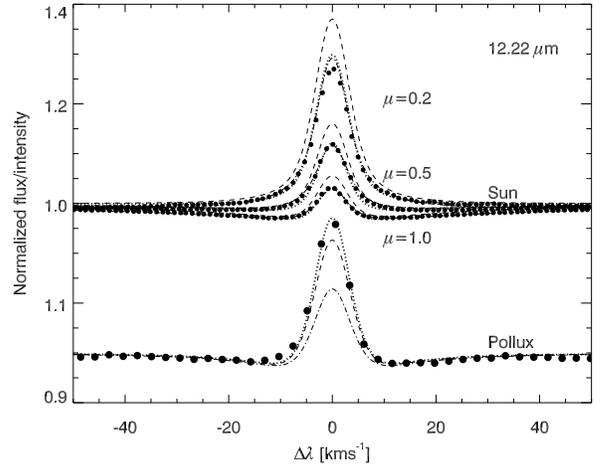}}
                \caption{Observed and modeled line profiles in Pollux
                and for three positions on the solar disk (indicated in the figure) 
                for the 12.22 $\rm\mu$m line, plotted 
                on a velocity scale. Labels as in Fig.\ \ref{Fig:12.22um_ArAl}.}               
        \label{Fig:12.22um_PoSu}
\end{figure}

\subsection{Emission lines at 12 $\rm\mu$m}
\label{12um}

\begin{figure*}
        \sidecaption
                \includegraphics[width=10cm,angle=90]{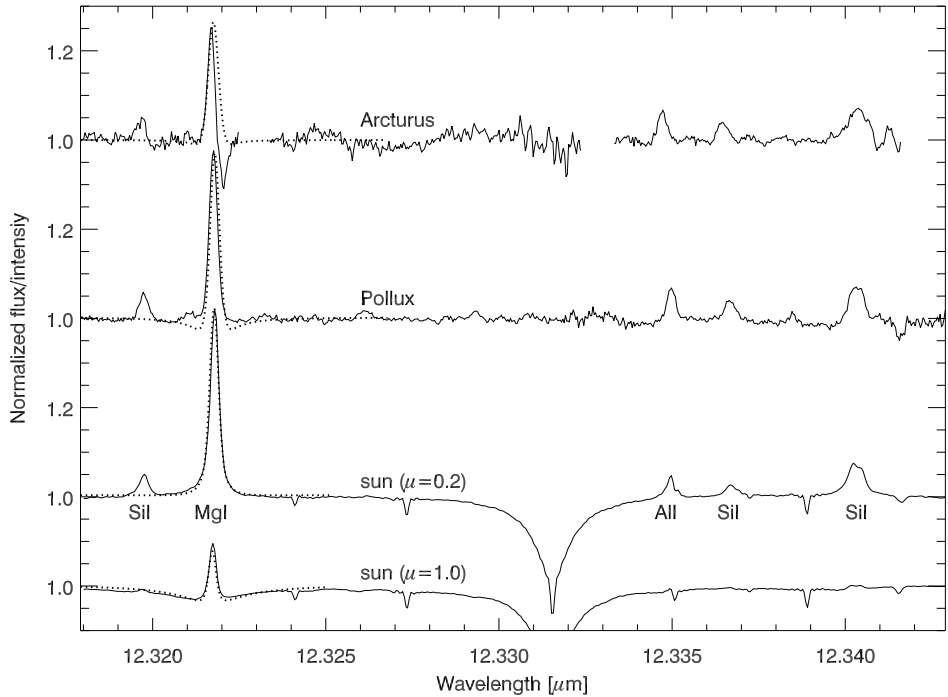}
                \caption{Observed spectra around the 12.3 $\rm\mu$m region, shifted to the
                solar frame, for Pollux, Arcturus, and for two positions 
                on the solar disk (indicated in the figure). Solar observations are from \citet{Brault83}. Models for 
                the Mg\,I 12.32 $\rm\mu$m line are labeled with dots, and use the 
                n=15;H model atom (see Fig.\ \ref{Fig:12.22um_ArAl}). The solar absorption
                features are telluric lines, and the missing parts in the 
                Arcturus spectrum are due to gaps between spectral orders. Visible emission
                lines as indicated in the figure. The Arcturus absorption line at 12.32 $\rm\mu$m is 
                a water vapor line.}              
        \label{Fig:12um_obs}
\end{figure*} 

We plot observed and computed line profiles for the 
Mg\,I 12 $\rm\mu$m lines in the Sun, Pollux, Arcturus,
and Aldebaran in Figs.\ \ref{Fig:12.22um_ArAl},
\ref{Fig:12.22um_PoSu} and \ref{Fig:12um_obs}.
In Arcturus and Aldebaran, the 12.32 $\rm \mu$m line is blended with 
a water vapor absorption line (see \citet{Ryde06} for an identification), 
which in the latter star is so influential 
that we choose to exclude the Aldebaran 12.32 $\rm\mu$m line from the analysis.  
The observed emission lines from the K giants are stronger than the solar lines. 
In addition to the Mg\,I lines, we also identify emission lines from 
Si\,I, Al\,I, and Ca\,I in the observed 
spectra of Pollux and Arcturus (three of the lines are displayed 
in Fig.\ \ref{Fig:12um_obs}), 
all identified as $n=7 \rightarrow 6$ transitions with high orbital 
quantum numbers. The flux maximum in the observed and normalized 
spectra (and the FWHM for the Mg\,I lines) are given in Tables \ref{Tab:12um_mg} 
and \ref{Tab:12um}. More line data can be found in \citet{Chang83} 
and \citet{Chang84}. We have also added 
results from modeled Mg\,I flux profiles for 
the Sun (from a disk integration over
solar intensity profiles in a model that reproduces 
the observations) in Table \ref{Tab:12um_mg}, 
to enable a fair comparison between solar and stellar observations.
This illustrates that the K giants have stronger emission than 
the Sun.\\ 
The line-center average depth of formation in the modeled Mg\,I 12 $\rm\mu$m 
lines is, for the Sun and Pollux, in atmospheric layers slightly below 
$\log \tau_{500} \sim -3$, with the weaker 12.22 $\rm\mu$m line shifted 
approximately 0.2 dex toward the inner photosphere. 
In Arcturus and Aldebaran, the line formation takes place deeper 
inside the atmosphere. The average depth of formation for the line-center in
the 12.22 $\rm\mu$m line in Arcturus and Aldebaran is $\log \tau_{500}=-1.8$
and $-1.6$ respectively. 
This is partly because of the lower amount of $H_{\rm ff}^{-}$ opacity in 
these atmospheres (due mainly to lower electron abundances), 
which shifts the continuum formation to about $\log \tau_{500} \sim -0.8$, 
as compared with $\log \tau_{500} \sim -1.2$ in Pollux.\\
The extension of the model atom has a significant impact on 
the synthetic line spectra, with computed intensity/flux profiles being
much stronger when using the enlarged model atom. 
We note the failure of the smaller atom to reproduce the observed 
emission for the K giants, whereas it provides a good match for the solar lines, 
in agreement with previous studies.  
The larger model atom without collisions with neutral hydrogen 
predicts emission lines well below the observed level for Arcturus and Aldebaran, 
in contrast to the Sun where the modeled lines now are too strong.  
However, when including collisions with neutral hydrogen (as described in  
Sect.\ \ref{Hcoll}) the models reproduce the observed emission
in all cases. These different responses to the `added' 
collisions demonstrate the complexity of the Rydberg state 
deexcitation ladder, and are further discussed in Sect.\ \ref{Hcoll_disc}.\\
Our models predict narrow absorption troughs in the Pollux lines, 
only matched by observations in the red wing of the 12.22 $\rm \mu$m line. 
However, due to uncertainties in the observed normalized spectra 
imposed by, e.g., the continuum setting, we are not able 
to draw any firm conclusions from the absence of absorption troughs. 
A discussion about shifts in the solar absorption troughs, visible in Fig.\ \ref{Fig:12um_obs}, 
can be found in, e.g., \citet{Chang94}. We also note how the 
line-wings in Arcturus and Aldebaran are too broad to be fitted by an 
isotropic Gaussian, and require a radial-tangential macroturbulence 
(see Sect.\ \ref{atmospheres}).

\subsection{Mg\,I emission lines at 18 $\rm\mu$m}
\label{18um}

\begin{figure}
                \resizebox{\hsize}{!}{\includegraphics[angle=90]{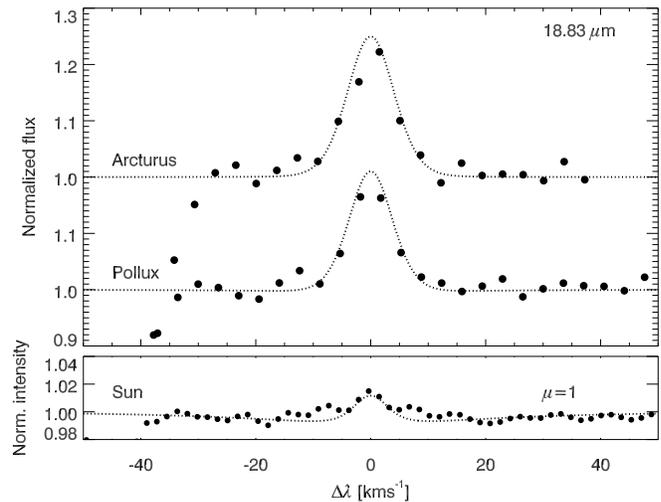}}
                \caption{Observed and modeled line profiles for the 18.83 $\rm\mu$m
                line in Pollux, Arcturus, and at the solar disk-center. 
                Labels as in Fig.\ \ref{Fig:12.22um_ArAl}. The feature to the left 
                in the figure is an OH absorption line.
                Note the scale difference between the upper and lower ordinate.}           
        \label{Fig:18um}
\end{figure}

We also observed the $8h \rightarrow 7g$ Mg\,I transition 
at 18.83 $\rm \mu$m in Pollux and Arcturus, and present 
here the first stellar observations of this line. 
The emission is high here as well, see 
Fig.\ \ref{Fig:18um} and Table \ref{Tab:12um_mg}.
Our synthetic line spectra reproduce the observed emission 
also for this line, which suggest that our model atom accounts for the 
Rydberg state deexcitation ladder in an accurate way. For comparison reasons, we 
display also a solar disk-center intensity profile\footnote{Observations 
from the Kitt Peak solar atlas}. The observed solar line
feature is barely visible, which further illustrates the stronger 
emission from K giants.\\  
The difference between departure coefficients in $n=8 \rightarrow 7$ transitions is of
similar magnitude as that between 7 and 6, 
causing comparable emission line strengths. 
The continuum formation is shifted 
about 0.3 dex outwards when compared 
to the spectral region around 
12 $\rm\mu$m (the $H^{-}_{\rm ff}$ 
opacity increases) but the average height of 
formation for the line-center in the 18.83 $\rm\mu$m line 
in Pollux is located at $\log \tau_{500} \sim -2.5$, 
slightly further in than the 12 $\rm\mu$m lines.
This is because the line is merely the 
third strongest $8l \rightarrow 7l'$ transition with $\Delta l = -1$. 
We thus predict that the next two Rydberg 
transitions in the chain ($8i \rightarrow 7h$ located at 
18.99 $\rm \mu$m and $8k \rightarrow 7i$ at 19.03 $\rm \mu$m) 
should appear even stronger, however these were not 
covered in our observational setup.   

\begin{table}
      \caption{Summary of observed magnesium emission line properties.} 
         \label{Tab:12um_mg}
         \centering
         \begin{tabular}{lcccccc}
            	\hline\hline	 
            	& \multicolumn{3}{c}{Normalized} & \multicolumn{3}{c}{FWHM}  \\
            	& \multicolumn{3}{c}{flux maximum} & \multicolumn{3}{c}{[$\rm km\,s^{-1}$]}  \\
            	Star, \ \ \ line [$\rm \mu$m] &12.22&12.32&18.83&12.22&12.32&18.83 \\
            	\hline
              Pollux     &        1.28&1.37&1.20            & 6.8&7.1&8.2  \\
              Arcturus	&	  1.24&$^a$&1.23	&  8.0 &$^a$&9.7 \\
              Aldebaran &   1.16&$^a$& - &  8.0 &$^a$& - \\
              Sun$^b$   &   1.09&1.15&- & 5.8&6.1 &- \\   	
              \hline 
              \multicolumn{7}{l}{}                                   \\                              
              \multicolumn{7}{l}{$^a$ Emission line blended with $\rm H_2 O$ absorption line.}     \\
              \multicolumn{7}{l}{$^b$ Modeled solar flux profiles (see text).}
         \end{tabular}
\end{table}

\begin{table}
      \caption{Summary of observed emission line properties around 12 $\rm\mu$m (other elements than 
magnesium) for Pollux and Arcturus.} 
         \label{Tab:12um}
         \centering
         \begin{tabular}{lcccc}
            	\hline\hline	
            	Element$^a$  & \multicolumn{2}{c}{Wavenumber$^b$} 
            	 & \multicolumn{2}{c}{Normalized} \\
            	 &  \multicolumn{2}{c}{Wavelength}  & 
            	 \multicolumn{2}{c}{flux maximum}  \\
            	 &[$\rm cm^{-1}$]&[$\rm\mu$m]&Pollux&Arcturus\\
            	\hline
              Si\,I     &	 810.360&12.340	& 1.07&1.08     \\  
              Si\,I	&	 810.591&12.337	& 1.04&1.04      \\
              Al\,I 	&	 810.704&12.335	& 1.07&1.06	\\
              Si\,I	&        811.709&12.320	& 1.06&1.05	   \\    
              Si\,I	&	 813.380&12.294	& 1.06&1.07	   \\ 
              Si\,I$^c$ &        814.273&12.281 & 1.04& -          \\
              Ca\,I$^d$	&	 814.969&12.270	& 1.05&1.02 	   \\    
              Al\,I$^c$ &        815.375&12.264 & 1.03& -          \\
              Si\,I$^c$ &        815.979&12.255 & 1.03& -          \\    
            	\hline 
           \multicolumn{5}{l}{}                                   \\ 
           \multicolumn{5}{l}{$^a$ Line identifications based on} \\ 
           \multicolumn{5}{l}{\ \ \ \citet{Chang83} and \citet{Chang84}.} \\
           \multicolumn{5}{l}{$^b$ From the solar observations by \citet{Brault83}.}   \\                           
           \multicolumn{5}{l}{$^c$ Emission line blended with OH absorption line.}     \\
           \multicolumn{5}{l}{$^d$ Line identification in \citet{Chang83}}             \\
           \multicolumn{5}{l}{\ \ \ stated as `suspicious'.}   \\

         \end{tabular}
\end{table}


\section{Discussion and conclusions}
\label{Discussion}

The model atom extension and the introduction of collisions
with neutral hydrogen remove the discrepancy between observed 
and modeled emission in our study.
\textit{Before} the model atom was extended, we 
undertook a number of tests to investigate the large 
discrepancies between observations and models.
We therefore start our discussion by a short summary of 
these, with the purpose to simplify future work.\\ 
The lines are sensitive to 
the photoionization rates as these affect 
the recombination to the Rydberg levels. We thus made an
\textit{ad-hoc} increase in the photospheric MARCS 
radiation field until the modeled Mg\,I emission 
lines matched the observed, but found that 
the required additional mean intensities 
gave rise to surface fluxes that by far 
exceeded observed ones. We included a chromospheric 
temperature-rise in Arcturus and found it to 
have a negligible impact, with the lines forming in atmospheric 
layers below the temperature minimum. 
We computed MARCS models in spherical geometry and employed 
the spherical version of MULTI, S-MULTI \citep{Harper94}, but 
differences from plane-parallel models were small.
No attempt to analyze influences from atmospheric inhomogeneities 
has been made in this work. A discussion about 
how granulation affects the solar lines can be found in \citet{Rutten94}.   

\subsection{The model atom extension}
\label{Extension}

All Mg\,I departure coefficients are shifted 
upwards (increased) by the extension of the model atom, 
but the upward shift is more pronounced in the higher energy levels.
It is this change in the departure 
coefficient \textit{ratio} in the line-forming 
regions that is sufficient to cause a significant 
change in results, i.e., higher emission 
peaks for the larger atom. The enhanced collisional coupling in the uppermost 
Mg\,I levels and to the Mg\,II ground state strengthens the cascading 
process in rather the same manner as the model atom with top-levels $n=9$ did when compared to 
one reaching only $n=7$ (see Sect. \ref{Enlargement}). Qualitatively, more recombinations enter at the top-levels,
channel down through transitions that take part of the Rydberg ladder (see Fig.\ \ref{Fig:terms})
and set up larger departure coefficient differences between adjacent levels. 
The extension thus has a significant impact on the 
mid-infrared emission lines, whereas the overall character 
of the Mg\,I statistical equilibrium remains.\\ 
To include the energy levels with $n \ge 10$ seems especially pertinent when applying the model atom
on diluted stellar atmospheres with low surface gravities (as shown by the large 
differences in the modeled line profiles of the K giants). These
are more influenced by radiative transitions, and thus the extension ensures 
that the Rydberg level replenishment from the ion state is properly accounted 
for by including top-levels that are fully dominated by collisions. 

\subsection{Effects from extra collisions}
\label{Hcoll_disc}

Higher rates of collisional excitation and ionization 
affect the 12 $\rm\mu$m lines in the 
implemented stellar model atmospheres differently. 
Figs.\ \ref{Fig:12.22um_ArAl} and \ref{Fig:12.22um_PoSu} show how (by the introduction of 
collisions with neutral hydrogen) the Mg\,I 12 $\rm\mu$m emission is reduced 
in the Sun, increased in Arcturus and Aldebaran and almost unchanged in Pollux.
Apparently, a homogeneous increase of collisional rates actually results in stronger emission 
in the low surface gravity atmospheres of Arcturus and Aldebaran 
(which one would perhaps not expect since, generally, collisions 
act to thermalize lines toward LTE).\\ 
We have analyzed this result by computing additional models for the Sun and Arcturus, 
where we included radiatively allowed bound-bound collisions with neutral hydrogen for 
1) only the three transitions $7i \rightarrow 6h$, $7h \rightarrow 6g$, and $7g \rightarrow 6f$
and 2) only transitions with $\Delta n,\Delta l = -1$ and $n \geq 8$ 
and $l \geq 4$. The first test imposes stronger collisional coupling only 
in the $n = 7 \rightarrow 6$ transitions themselves, whereas 
the second serves to strengthen the collisional bound-bound 
coupling in transitions in the uppermost levels that
take part of the Rydberg ladder (see Fig.\ \ref{Fig:terms}), while maintaining all other rates.\\ 
We quantify the results of this experiment by the normalized 
flux maximum in the 12.22 $\rm\mu$m line. 
In the first case, both models give lower emission, as expected from the 
direct thermalizing effect on the $n = 7 \rightarrow 6$ transitions. The modeled 
flux maximum above the continuum decreased by $\sim$\,30\,\% 
and $\sim$\,25\,\% for the Sun and Arcturus respectively, as 
compared with the model excluding collisions with neutral hydrogen. 
For the second case, however, the emission increased with a similar percentage 
in the solar model, whereas in Arcturus, the modeled flux maximum 
doubled its value. Thus, in a star like Arcturus the enhancement effect 
from highly excited lines (see Sect.\ \ref{Departure}) dominates 
the reduction effect from the higher collisional rates in the line transitions themselves, 
so that when introducing collisions with neutral hydrogen 
homogeneously throughout the model atom, the outcome is an emission increase 
(as seen in Fig.\ \ref{Fig:12.22um_ArAl}).      
We can understand this by noting that, e.g., the ratio 
between collisional and radiative deexcitation rates 
for the $8i \rightarrow 7h$ transition (supplying the 12.22 $\rm\mu$m line) 
is $C_{ji}/R_{ji}$\,$\sim$\,15.0 (Sun) and 
$\sim$\,0.2 (Arcturus) in typical line-forming layers when collisions with neutral 
hydrogen are excluded. In principle, this means that the contribution from hydrogen is needed 
in the giants to ensure an efficient Rydberg ladder.

\subsection{Observations of Rydberg emission lines around 12 $\rm\mu$m}
\label{Observations_disc}

The observed emission-line flux spectra for Pollux and Arcturus in the 12 $\rm\mu$m region closely
resembles the solar limb intensity spectrum, whereas the solar 
disk-center spectrum lacks emission features from other elements 
than magnesium (see Fig.\ \ref{Fig:12um_obs}). 
It is evident that strong emission features from the K giants, as compared with 
solar-type dwarfs, appear for more metallic Rydberg lines than 
magnesium, and the observability of different elements 
so far follows the same pattern as in the Sun.  
Future observations will tell if this observed 
trend remains for a larger sample. Model tests with K dwarfs 
indeed predict lower Mg\,I emission for dwarfs than for giants also 
within the same spectral class (supported 
as well by observations of the magnetically active K dwarf 
$\epsilon$ Eridani, Richter et al., in preparation).\\
For cooler K giants, absorption from water vapor starts 
to influence the 12 $\rm\mu$m spectrum and we have 
detected a blend at 12.32 $\rm\mu$m in Arcturus and Aldebaran.
Observations from this spectral region in the yet cooler M supergiant 
Betelguese \citep{Ryde06} reveal no emission lines above the noise 
level, which is consistent with our modeling of the Mg\,I lines 
in this star (using the same stellar parameters as in \citeauthor{Ryde06}).

\subsection{Comparison with other studies}

\citet{Uitenbroek96} modeled the Mg\,I 12.32 $\rm\mu$m line in 
Arcturus and concluded that the computed line was, roughly, 
half as strong as the observed. When using the 'C92' 
model, we find an even larger discrepancy for the 
12.22 $\rm\mu$m line (see Fig.\ \ref{Fig:12.22um_ArAl}).
As already discussed, these longstanding discrepancies between 
observations and models for K giants are removed when using our new model atom. 
We note also that \citeauthor{Uitenbroek96} did 
not detect the water vapor absorption line, which is 
blended with the Mg\,I 12.32 $\rm\mu$m line in our Fig. \ref{Fig:12um_obs}. 
The extension of the model atom changes the results in the 
solar case as well (as compared with C92 and \citeauthor{Ryde04} 2004), 
placing the Mg\,I lines in higher emission, but the
former results are recovered by the introduction of collisions 
with neutral hydrogen.\\  
We have settled here with the rather questioned, albeit standard, 
Drawin recipe for collisions with neutral hydrogen while 
we await results from more rigorous quantum mechanical calculations. 
We have shown the influence from these collisions
on the formation of the mid-infrared Mg\,I emission lines, however 
we stress that it has not been an aim of this investigation to put detailed
empirical constraints on their efficiency. Such a task would require a larger set of 
lines, including also other wavelength regions. Nevertheless, we may still 
compare our adopted scaling factor to the Drawin formula 
for Mg\,I collisions with neutral hydrogen, 
$S_{\rm H}=10^{-3}$, with other values from the literature.
Mentioned in Sect.\ \ref{Atom} was the exponential decrease 
resulting in $S_{\rm H}=3 \times 10^{-10}$ for the 12 $\rm\mu$m transitions
\citep{Zhao98}, a model which was later abandoned  
by the same group in favor of a constant $S_{\rm H}=0.05$ \citep{Gehren04}, 
inferred only from optical lines. 
In a recent non-LTE abundance study of magnesium in
metal-poor stars \citep{Mashonkina08}, $S_{\rm H}=0.1$ is used.
The value we find based on the mid-infrared lines is more than one 
order of magnitude lower than the values obtained from these two optical studies.
Our model atom has not been applied to optical lines, however such a 
combined study should be given high priority in future work. 
We thus conclude that the mid-infrared emission lines from near-by giant stars 
may be suitable diagnostics for testing atomic input data in future 
non-LTE analyses.\\
Finally, as we are now able to model and explain the observed emission lines for 
both dwarfs and giants, diagnostic applications regarding stellar disk-averaged 
magnetic fields are possible.

\begin{acknowledgements}

We thank M.\ Carlsson for providing the data and 
procedures from his model atom, and
J.\ Puls, K.\ Lind \& K.\ Butler for careful readings 
and valuable suggestions to the manuscript.\\ 
JS gratefully acknowledges a grant from the International Max-Planck 
Research School of Astrophysics (IMPRS), Garching. NR is Royal Swedish 
Academy of Sciences Research Fellow supported by a grant from the
Knut and Alice Wallenberg Foundation. GH.'s contribution was 
supported by NASA ADP grant NNG 04GD33G issued through the 
Office of Space Science and the NSF US-Sweden Cooperative 
Research Program grant INT-0318835 to the University of Colorado. 
We are also grateful for financial support from The Swedish Foundation 
for International Cooperation in Research and Higher Education (STINT), 
grant IG 2004-2074. MR and AK acknowledge the support from the NSF grant AST-0708074.\\ 
Observations with TEXES were supported by NSF grant AST-0607312 and we thank the other 
members of the TEXES team for their assistance. We also acknowledge the excellent support of 
Gemini and IRTF day and nighttime staff who help make TEXES observations a success.
NSO/Kitt Peak FTS data used here were produced by NSF/NOAO.

\end{acknowledgements}

\bibliographystyle{aa}
\bibliography{references_Mg}

\end{document}